# Site Reliability Engineering: Application of Item Response Theory to Application Deployment Practices and Controls

**Kiran Mahesh ND**
NCR Corporation, Hyderabad (email: {Mahesh.nalamdurga} @ncr.com)

**ABSTRACT** Reliability of an application or solution in production environment is one of the fundamental features where every SRE team is critically focused upon. At the same time achieving extreme reliability comes with the cost which include but not limited to slow pace of new feature deployments, operations cost and opportunity cost. One such earlier effort in giving an objective metric to strike the fine balance between acceptable reliability and product velocity is error budget and its associated policy. There are also contemporary deployment guidelines and controls per organization to ascertain the reliability of an application deployment version into customer facing or production environments. This work proposes new objective metrics called *Application Deployment Score* estimated using dichotomous Item Response Theory model. This score is used to assess the improvement trend of each application version deployed into customer facing environment, identify the improvement scope for each application deployment in each area of deployment guidelines and controls, adjust the error budget i.e. *soft error budget* of a interdependent application in application mesh by giving soft collective responsibility and finally defines a new metric called *deployment index* which helps to assess the effectiveness of these contemporary deployment guidelines and controls in upholding the agreed SLOs of the application in customer facing environments. This study opens a new field of research in developing new underlying latent indexes (i.e. new objective metrics) in SRE and DevOps space.

## I. Introduction

In an application platform framework various services works together in synergy to support the solutions that fuel the modern missions of many software led organizations. Reliability of these services or applications are of paramount importance. Software reliability is defined as the probability of failure free software operation for a specified period of time [6]. Site Reliability Engineering principles, practices and incentives rotate around software engineering way of handling the reliability of an application or service in customer facing environment [13]. There are organizational level established application deployment guidelines and controls (checklists) to assess the risk of an application deployment in customer facing environment. There is a need to objectively measure the adherence of respective application deployment to these checklist items. One such attempt is the introduction of new metric called Application Deployment Score (ADS) in this work. This score is calculated using the application of Item Response Theory (IRT) which had been applied in various fields right from psychometry to education. Ability of an application to adhere to established controls is a latent trait and has been incorporated to calculate the application deployment score.

Another approach to manage application or service reliability is by managing the risk associated with new feature deployments. An objective metric called Error budget [1] is introduced to strike this fine balance between application or service reliability and new feature release velocity. This budget takes into consideration the collective responsibility of upholding the agreed SLAs. This works introduces another metric called Soft Error Budget which sits in the space generated by two extreme ways of calculating the error budgets in an interdependent service mesh. This soft mode of calculating the error budget takes in to consideration the calculated application deployment score of respective deployment version of an application in customer facing environment and penalizes for non-adherence of this application version to established controls.

So far its being discussed on how important an application deployment to adhere to established guidelines and controls but it is equally important to objectively measure the effectiveness of established contemporary deployment guidelines and controls to uphold the agreed SLAs of services deployed in customer facing environments. This work proposes another new metric called Deployment Index which is calculated by taking correlation between Application deployment score for each deployment of respective application or service and Average achieved SLO [2] for that deployment in the customer facing environment.

## II. Background

This section introduces Error Budget, SLOs and Dichotomous IRT model

### A. Service Level Objectives (SLOs)

SLO is defined as a threshold value or a range of values within which service level indicator (SLI) of an application resides [2]

$SLI \leq$ Agreed SLO or

SLO Lower bound $\leq SLI \leq$ SLO Upper bound

Service level indicator is an objective measure of current level of a service (application) in customer facing environment. Few of the heavily used such indicators include Response time latency, error rate and service (or application) availability. There are instances where SRE teams use composite SLOs which are weighted average of two or more SLOs in their highly complex environments.

### B. Error Budget

To have a fine balance between application reliability in customer facing environments and rapid innovation from product teams, we use an objective metric called Error Budget [1]. In short it defines how much unreliable (or downtime) a service is allowed in a giver customer facing environment.

$$\text{Error Budget} = 1 - \text{SLO} \qquad (1)$$

If agreed Availability SLO for a service is 'four nines' i.e. 99.99% then allowed Error budget for that service is 0.01% (allowed downtime in customer environment) which boils down to 4.32 minutes of downtime in one month (0.01% * 30 days = 0.003days = 0.072 hours = 4.32 minutes)

### C. Dichotomous Item Response Theory model

Item Response Theory or latent response theory is a consolidation of mathematical models which helps to measure the underlying variable of interest based on its response to an instrument (i.e. a test or a checklist) [3]. Here variable of interest is something that can be intuitively understood (e.g. emotional intelligence of a student) but cannot be measured directly (like weight of a person). For the scope of this current work, we refer the ability of deployment of an application to adhere to established guidelines and controls as the underlying variable of interest or latent trait.

Unidimensional IRT considers the presence of such single latent trait (i.e. ability of deployment of an application to adhere to established guidelines and controls) that will be measured with mathematical models throughout the instrument (i.e. a consolidated checklist)

Items (each item in a checklist) of an instrument (i.e. checklist) with two possible outcomes i.e. whether an application deployment adheres to the item or not (i.e. 1 or 0; Pass or Fail) are referred as dichotomous items and IRT mathematical models associated with these datasets are called dichotomous item response theory models

IRT mathematical models:

$$P(\theta) = \frac{1}{1+e^{-L}} \qquad (2)$$

One parameter Logistic model (1 PL) or Rasch model: L = Ø-b

Two parameter Logistic model (2 PL): L = a(Ø-b)

a = item or checklist discrimination

b = item or checklist difficulty

Ø = the ability or underlying latent variable (i.e. ability of deployment of an application to adhere to established guidelines and controls

P(Ø) = Probability of correct response or adherence of application deployment to respective checklist item

Another extension of these models called 3 PL which include guessing parameter into consideration.

Assumptions of IRT model include unidimensional nature of latent trait, local independence of items, monotonicity and group invariance.

## III. Related Work

This section discusses work related application of Item Response Theory [3] to software Reliability Engineering practices, error budgets. Objective metrics in application reliability engineering is predominantly found in Google SRE handbooks [1]. Few attempts include applying IRT to estimate the usability of e-commerce web application [5]. Framework for error budget and managing the risks [7] [12] has be dealt in varying degree which include application of agile principles [8], common mistakes around SLO calculations [9] and its alerting setting [10] and extending the framework with addition of few additional dimensions like security and feature freshness [11]. Assessing the effectiveness of deployment guidelines and controls in deploying reliable applications to customer facing environments are limited.

## IV. Methodology

This section first defines the new objective metric called *Application Deployment Score* and its application in giving a progress report for application deployments in customer facing environments. Secondly it explains the novel concept of *soft error budget (adjusted error budget)* of an application and finally a new metric called *deployment index* to assess the effectiveness of deployment guidelines and controls in upholding the agreed SLOs of the application in customer facing environments.

### A. *Application Deployment Score*

This score for each application deployment is also referred as true score is summation of individual item's probability of success (or probability of adherence of application deployment to respective checklist item).

$$\text{Application Deployment Score (ADS) for application deployment j} = \sum_{i=1}^{n} P_i(\theta_j) \qquad (3)$$

i = item in application deployment response vector

j = application deployment

$P(\emptyset)$ = Probability of correct response or probability of adherence of application deployment to respective checklist item from equation (2)

---

Algorithm 1: Calculating ADS

---

*Step 1:* Prepare the response vectors of items in deployment guidelines and controls for each application deployment into customer facing environment

*Step 2:* Prepare the response matrix which consists of response vectors of all the application deployments for an agreed time window. This matrix forms the basis for training the chosen mathematical model (2 PL in this case)

*Step 3:* Fit the 2 PL Dichotomous IRT model with the processed dataset (in *step 2*) using variants of maximum likelihood estimate (MLE) which is being implemented in various python / R packages (ltm package in R is used in this work)

*Step 4:* Learn the item (checklist) parameters b, a for all the items in training dataset and the ability or underlying latent variable $\emptyset$ for respective application deployments in the training dataset

*Step 5:* Calculate the application deployment score for each application deployment from the learned parameters and its respective response vector from equation (3)

*Step 6:* Prepare the report for all application deployments highlighting all its scores

*Step 7:* Fit the ability or underlying latent variable for new application deployment using the available response vector to the deployment checklist and controls

---

Calculated Application Deployment Score (ADS) helps is assessing the respective deployment of an application in customer facing environments and gives the true indicator of how well the respective deployment of an application adheres to the established contemporary deployment guidelines and controls of an organization. This score helps in understanding how an application is improving or degrading in adhering to the established deployment guidelines and controls of an organization. It gives a trend of score over the deployment calendar and helps recommend the areas of improvement in established deployment guidelines. ADS form the basis for a recommendation model to recommend areas of improvement for each application to make it highly compliant to established deployment guidelines and controls. It also forms as a progress report for organizational level application deployments and helps assess the maturity of our deployment process.

### B. Soft Error Budget

In modern loosely coupled service architecture that helps ease service integrations and discoveries across platforms, organizations combine these API ready individual services to create innovative customer solutions or products. In this journey one service or application may depend on another service in solution flow and impacts the overall reliability of that solution or product to customers. Figure (1) gives a glimpse of this:

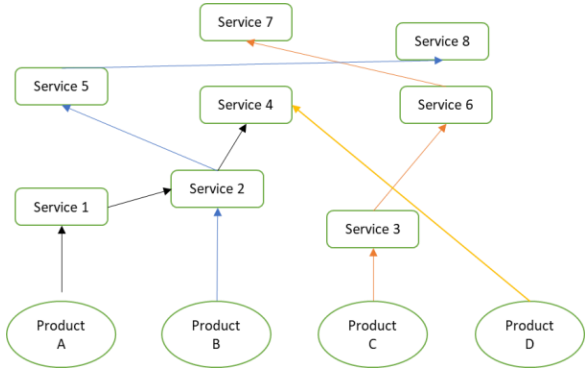

Figure 1

Below are the assumptions to services that participate in a product or solution offering to customers:

a) Service independence i.e. true micro service
b) Availability of monitoring solution to capture availability at each service level independently and hence the calculation of error budget of each service independently excluding dependency downtimes

SLAs for one product is different from that of other product and so are their respective SLOs. Services at different hierarchical level of a given product shares the collective responsibility in calculating the available error budget for that product releases. These individual services are guided by respective team's feature deployment velocity.

Having a notion of collective responsibility to calculate the error budget in a dependent environment is acceptable but in modern service mesh, we need to adjust the error budgets by relaxing this notion to calculate the true error budget for each service and improves the new feature velocity of respective interdependent service (e.g. service 2 for product A in Figure (1)).

We have two extremes in calculating error budget:

a) Measure the available error budget at product or solution level and apply that error budget equally to all the dependent services in that solution irrespective of whether that service contribute to the overall solution availability. This holds the notion of taking the collective responsibility for product or solution SLO. This is also referred as top down approach
b) Treat each service independent of each other in that solution flow and calculate the error budget independently to guide new feature deployments to that service. This is also referred as bottom up approach

This work introduces a novel approach in adjusting the error budget for each service called soft error budget which sits in the space generated by two extreme error budgets mentioned above. This takes into consideration the true risk associated with other interdependent services in the product or solution and adjusts the Error budget to make it accountable to the overall solution it integrates. Equation (4) defines the Soft Error Budget:

$$Soft\ Error\ Budget_i = Error\ Budget_i - \sum_{j=1, j \neq i}^{n} \left(1 - \frac{ADS_j}{Total\ Items}\right) * Unavailability_j \quad (4)$$

i = service for which error budget is adjusted

j = other services available in the given product or solution

n = total number of services participating in a product or solution

*Total Items* = Total checklist items that are considered while calculating the ADS for services

$Error\ Budget_i$ = Service error budget measured independently excluding dependency downtimes

$\left(1 - \frac{ADS_j}{Total\ Items}\right)$ = True Risk factor for service j (i.e. other available service in solution or product)

*ADS* = Application Deployment Score

*Unavailability* = application downtime, error rate etc. which is measured independently for each service excluding dependency downtimes

| Algorithm 2: Calculate soft error budget |
|---|

*Step 1:* Define the SLOs for all the customer facing products or solutions in service mesh. Services in a product or solution will adhere the respective SLOs and hence the allocated error budget.

*Step 2:* Calculate the Application deployment score (ADS) of all service deployments in the environment from equation (3)

*Step 3:* For a given time frame, calculate the available error budget of each service independently using equation (1)

*Step 4:* Calculate the soft error budget using the equation (4)

If a given service is participating in one or more product or solution offerings, this work recommends taking minimum of the soft error budgets of the service obtained by applying respective product or solution SLOs.

### C. Deployment Index

To assess the effectiveness of these contemporary deployment guidelines and controls in upholding the agreed SLOs of the application in customer facing environments, this work proposes a new metric called Deployment index defined in equation (5). This index is calculated by taking correlation between Application deployment score for each deployment of respective application or service and Average achieved SLO for that deployment in the environment.

$$\text{Deployment Index}_{application} = \text{Corr}(\text{ADS}_{deployment}, \text{Avg Achieved SLO}_{deployment}) \quad (5)$$

ADS = Application Deployment Score for each deployment version of the application or service

Avg Achieved SLO = Average SLO achieved for respective version of application over the time (Normalized on duration of version available in customer facing environment)

This index forms the basis of understanding how effective the established contemporary guidelines and controls in maintaining the agreed SLO of that application or service.

## V.     Experiment

In this section the proposed methodology of calculating Application Deployment Score is evaluated on the real-world dataset obtained from application deployments in retail platform. This section will not cover calculations of Soft Error Budget and Deployment Index which predominantly use calculated ADS for respective service due to limited or unavailability of data and remains as a proposed conceptual construct for the scope of this work.

Current dataset consists five limited deployment controls or checklist items which include two controls on monitoring / alerting tools and three on application security tools. These five controls or deployment checklist items validates whether respective application passes the respective tool checklist (given a score of 1) or failed to pass (given a score of 0).

Applied 2 PL Dichotomous IRT model [3] to estimate the checklist item parameters and respective application deployment score (from the obtained ability parameter i.e. latent trait).

Accuracy of the calculated latent trait(ability) parameter is estimated using the item characteristic curve and predominantly test characteristic curve. Higher the information, higher is the estimation accuracy of the calculated parameters.

## A. Results and Discussion

ltm package in R [4] is used to implement the proposed methodology to calculate Application Deployment Score (ADS) per application deployment. Table (1) depicts the Ability (i.e. latent trait) calculated for available application deployments and hence the ADS against five deployment controls.

| Applications | Control A (Monitoring Tool) | Control B (Alerting Tool) | Control C (Security Tool) | Control D (Security Tool) | Control E (Security Tool) | Ability (Latent Trait) | Total Score | Application Deployment Score (ADS) |
|---|---|---|---|---|---|---|---|---|
| App 1 | 0 | 1 | 0 | 0 | 0 | -0.6222363 | 1 | 1.099 |
| App 2 | 1 | 1 | 1 | 0 | 0 | 0.67737895 | 3 | 2.959 |
| App 3 | 1 | 1 | 0 | 0 | 1 | 0.01492273 | 3 | 2.257 |
| App 4 | 1 | 1 | 1 | 0 | 0 | 0.67737895 | 3 | 2.959 |
| App 5 | 1 | 1 | 1 | 0 | 1 | 0.78103791 | 4 | 3.015 |
| App 6 | 1 | 1 | 1 | 0 | 1 | 0.78103791 | 4 | 3.015 |
| App 7 | 1 | 1 | 0 | 0 | 0 | -0.0305044 | 2 | 2.186 |
| App 8 | 0 | 0 | 0 | 1 | 0 | -1.0790282 | 1 | 0.51 |
| App 9 | 1 | 1 | 1 | 0 | 0 | 0.67737895 | 3 | 2.959 |
| App 10 | 1 | 1 | 0 | 1 | 0 | -0.0656796 | 3 | 2.13 |
| App 11 | 1 | 1 | 1 | 0 | 0 | 0.67737895 | 3 | 2.959 |
| App 12 | 0 | 0 | 0 | 0 | 1 | -0.9605057 | 1 | 0.615 |
| App 13 | 0 | 0 | 0 | 0 | 0 | -1.0228723 | 0 | 0.556 |
| App 14 | 0 | 0 | 0 | 0 | 0 | -1.0228723 | 0 | 0.556 |
| App 15 | 1 | 0 | 0 | 1 | 0 | -0.3957715 | 2 | 1.533 |
| App 16 | 0 | 0 | 0 | 0 | 0 | -1.0228723 | 0 | 0.556 |
| App 17 | 0 | 0 | 0 | 0 | 0 | -1.0228723 | 0 | 0.556 |
| App 18 | 0 | 1 | 0 | 0 | 0 | -0.6222363 | 1 | 1.099 |
| App 19 | 1 | 0 | 1 | 0 | 0 | 0.09721414 | 2 | 2.378 |
| App 20 | 1 | 1 | 1 | 0 | 0 | 0.67737895 | 3 | 2.959 |
| App 21 | 1 | 1 | 1 | 0 | 0 | 0.67737895 | 3 | 2.959 |
| App 22 | 1 | 1 | 0 | 0 | 0 | -0.0305044 | 2 | 2.186 |
| App 23 | 1 | 0 | 1 | 0 | 0 | 0.09721414 | 2 | 2.378 |
| App 24 | 1 | 1 | 0 | 0 | 0 | -0.0305044 | 2 | 2.186 |
| App 25 | 0 | 0 | 0 | 0 | 0 | -1.0228723 | 0 | 0.556 |
| App 26 | 1 | 1 | 1 | 0 | 0 | 0.67737895 | 3 | 2.959 |
| App 27 | 1 | 1 | 0 | 1 | 0 | -0.0656796 | 3 | 2.13 |
| App 28 | 1 | 1 | 1 | 0 | 0 | 0.67737895 | 3 | 2.959 |
| App 29 | 1 | 1 | 1 | 0 | 0 | 0.67737895 | 3 | 2.959 |
| App 30 | 0 | 1 | 1 | 0 | 0 | -0.2223332 | 2 | 1.861 |
| App 31 | 1 | 1 | 1 | 0 | 0 | 0.67737895 | 3 | 2.959 |
| App 32 | 1 | 1 | 1 | 0 | 0 | 0.67737895 | 3 | 2.959 |

| Applications | Control A (Monitoring Tool) | Control B (Alerting Tool) | Control C (Security Tool) | Control D (Security Tool) | Control E (Security Tool) | Ability (Latent Trait) | Total Score | Application Deployment Score (ADS) |
|---|---|---|---|---|---|---|---|---|
| App 33 | 1 | 0 | 1 | 0 | 0 | 0.09721414 | 2 | 2.378 |
| App 34 | 1 | 1 | 1 | 1 | 0 | 0.60340146 | 4 | 2.912 |
| App 35 | 1 | 0 | 0 | 0 | 0 | -0.3679826 | 1 | 1.587 |
| App 36 | 1 | 0 | 0 | 0 | 0 | -0.3679826 | 1 | 1.587 |
| App 37 | 1 | 1 | 0 | 0 | 0 | -0.0305044 | 2 | 2.186 |
| App 38 | 1 | 1 | 0 | 0 | 0 | -0.0305044 | 2 | 2.186 |
| App 39 | 1 | 1 | 1 | 0 | 0 | 0.67737895 | 3 | 2.959 |
| App 40 | 1 | 1 | 1 | 0 | 0 | 0.67737895 | 3 | 2.959 |
| App 41 | 0 | 0 | 0 | 0 | 0 | -1.0228723 | 0 | 0.556 |
| App 42 | 0 | 0 | 0 | 0 | 0 | -1.0228723 | 0 | 0.556 |
| App 43 | 0 | 0 | 0 | 0 | 0 | -1.0228723 | 0 | 0.556 |

*Table 1*

Figure (2) shows how the ADS changes with increasing ability (latent trait) of application deployment to adhere to the contemporary deployment guidelines and controls. Item characteristics curve in Figure (3) plots the probability of success of a deployment control w.r.t given ability (latent trait) of the respective application deployment. It shows the two technical properties of the deployment controls i.e. difficulty of a control to be adhered by the respective application deployment and discrimination of the deployment control. Deployment control D (one of the security tool) has an interesting property where the its probability of success decreases with increasing ability of an application to adhere to established controls. Accuracy of the calculated Application deployment scores hints for the consolidation of various deployment guidelines and controls into a common repository of items to effectively calibrate the assessment. Consolidation of the assessment (guidelines and controls) to happen across dimensions that include but not limited to monitoring, alerting, security, governance, high availability and deployment automation.

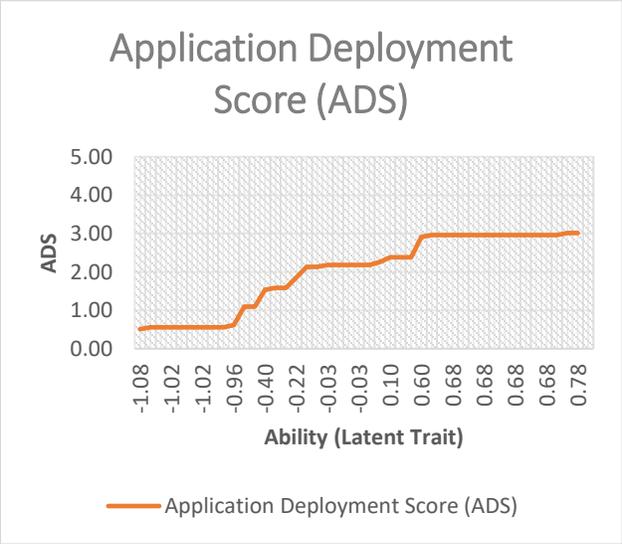

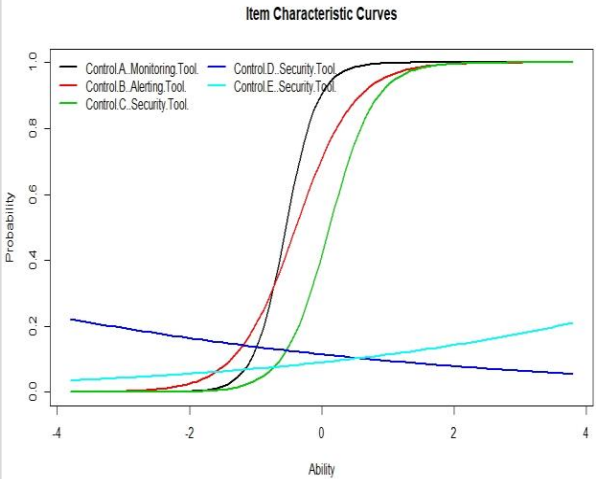

*Figure 2*

*Figure 3*

Accuracy of the calculated ability (latent trait) and hence the Application deployment score is estimated using Test information Function and Item information curves. Figure (4) and Figure (5) shows the presence of higher information around ability range [-1.0,0.5]. Higher the information, higher is the accuracy of the estimated parameter. Deployment controls to be carefully selected to spread across the entire ability range and with adequate amount of controls to estimate the application deployment scores. Ideally Information functions to be flat and horizontal.

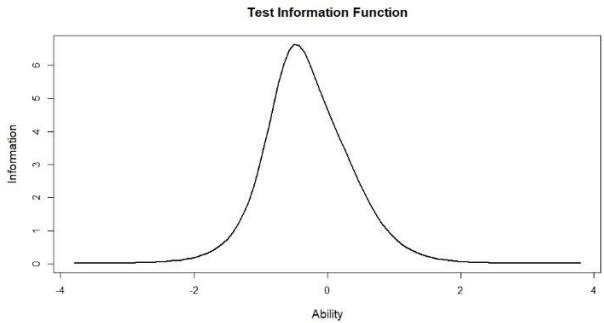
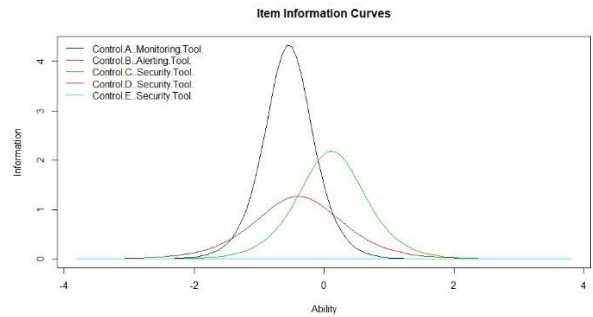

Figure 4                                                                                   Figure 5

## VI. Conclusion

With introduction of three new metrics around application reliability and organization's contemporary deployment guidelines, application deployment score forms the focal point. Application deployment score takes the base from the proven mathematical models of Item Response Theory. Accuracy of the calculated Application deployment scores hints for the consolidation of various deployment guidelines and controls into a common repository of items to effectively calibrate the assessment. Accuracy of the estimated parameters also supports the need of a robust instrument (consolidated checklist) spanning across various dimensions of assessing risk of an application deployment. Few of these dimensions include monitoring, alerting, high availability, security, governance and deployment automation. This work introduced objective metrics to adjust the error budget in an interdependent service mesh and provides the formulation to calculate the service level soft error budget in an interdependent customer facing environment. This metric takes the soft approach of adjusting the error budget. This work also introduced another metric to measure the effectiveness of contemporary organizational deployment guidelines and controls (checklists). This takes into consideration the agreed SLAs to customers and forms as a tool to adjust and refine these controls to serve the customers better. In short, this work gives a reference frame to measure the adherence of application deployments to organizational contemporary guidelines and controls, effectiveness of these controls to server the customers better and finally the soft notion of error budget to adjust the new feature velocity in interdependent platform applications.